# Magnetic anisotropy of FePt nanoparticles

Alamgir Kabir[a], Jun Hu[b], Volodymyr Turkowski[a], Ruqian Wu,[b] Robert Camley[a,c] and Talat S. Rahman[a]

[a] Department of Physics, University of Central Florida, Orlando, FL 32816
[b] Department of Physics and Astronomy, University of California, Irvine, CA 92697
[c] Department of Physics, University of Colorado, Colorado Springs, CO 80918

**ABSTRACT.** We carry out a systematic theoretical investigation of Magneto Crystalline Anisotropy (MCA) of $L1_0$ FePt clusters with alternating Fe and Pt planes along the (001) direction. The clusters studied contain 30 - 484 atoms. We calculate the structural relaxation and magnetic moment of each cluster by using *ab initio* spin-polarized density functional theory (DFT), and the MCA with both the self-consistent direct method and the torque method. We find the two methods give equivalent results for all the structures examined. We find that bipyramidal clusters whose central layer is Pt have higher MCA than their same-sized counterparts whose central layer is Fe. This results from the fact that the Pt atoms in such configurations are coordinated with more Fe atoms than in the latter. By thus participating in more instances of hybridization, they contribute higher orbital moments to the overall MCA of the unit. Our findings suggest that by properly tailoring the structure, one can avoid encapsulating the FePt $L1_0$ nanoparticles, as has been proposed earlier to protect a high and stable magnetic anisotropy. Additionally, using a simple model to capture the thermal behavior, we predict that a five-layered nanoparticle with approximately 700 atoms can be expected to be useful in magnetic recording applications at room temperature.

## I. Introduction

Understanding the physics of smaller structures can help in exploiting their useful properties. For example, the high surface-to-volume ratio and tailorable surface chemistry of metal nanoparticles have long been relied on in optimizing the activity and specificity of catalysts.[1] And small metal-particle arrays have been used to build single-electron devices.[2,3] Recently demand has arisen for magnetic particles with high anisotropic energy necessary for energy-harvesting technologies[4] as well as for ultra-high-density recording media.[5] Satisfaction of this demand requires development of metal thin-film media with smaller particles, more tightly-sized distributions, and optimized compositions.[6,7]

Since the mid-1930s Fe-Pt alloys of $L1_0$ phase have been known to exhibit high magnetocrystalline anisotropy.[8] Since among the various ferromagnetic metals and alloys FePt alloys show large perpendicular MCA (on the order of meV/atom[9]) and since, in nanoscale particles, they do not exhibit the superparamagnetism often characteristic of such small clusters,[10] it lends itself to magnetic applications requiring small-grained constructions. FePt alloys also have advantages over rare-earth transition-metal-based compound with high MCA, such as $Nd_2Fe_{14}B$ and $SmCo_5$ in that they are very ductile and chemically inert.[11] $L1_0$-based thin films and nanoparticles in general would seem to be promising candidates for ultra-high density magnetic storage media owing to their high corrosion resistance and excellent intrinsic magnetic properties.[12] But, in contrast to the fine grain of the $L1_0$ FePt systems, other conventional magnetic materials (Fe, Co, Ni and their alloys) would, through thermal fluctuation, within a very short time become superparamagnetic, losing any stored information. And given their high cost, bulk FePt-based permanent magnets can be used only for some especially delicate applications, as in magnetic micro-electromechanical systems (MEMS),[13] and in dentistry as attachment devices for retaining a dental cap in the cavity.[14]

The chemically ordered FePt $L1_0$ structure can be obtained by annealing from the fcc structure of FePt alloy or by deposition on substrate above the $L1_0$ ordering temperature.[15, 16] At high temperature an fcc solid solution of Fe-Pt is observed in the A1 phase; below 1573K (and down to 973K) alloys with a nearly equal number of Fe and Pt atoms (35-55% Pt) show order-disorder transition and the $L1_0$ ordered phase begins to form.[11] Though the $L1_0$ phase is typically obtained by heat treatment of A1 phase, it can also be produced by chemical synthesis of nanoparticles.[12] Deposition of alternate Fe and Pt monolayers can reduce the onset temperature of $L1_0$ phase.[17] Another way to obtain $L1_0$ phase experimentally is by annealing alternating multi-layers of Fe and Pt.[11] Stable FePt $L1_0$ nanoparticles have been prepared experimentally[18] both without any covering and with Al encapsulation.

The $L1_0$ structure has alternating Fe and Pt planes along the (001) direction, which is also the easy axis of magnetization, abandoning the cubic symmetry of the fcc system. In this type of layered magnetic system the MCA is mainly due to the contribution from the Pt (5d element) having large spin-orbit coupling while the Fe (3d element) provides the exchange splitting of the Pt sub-lattice.[19-21] It is well known that in the FePt system, the Pt atoms play an important role in magnetic anisotropy, because the hybridization of Fe orbitals cause spin polarization of Pt atoms, which in turn enhance the MCA owing to the relative strong SOC of the Pt atoms in comparison with that of the Fe atoms.

There have been several theoretical studies on MCA of nanoparticles. Cyrille *et al.*[22] calculated the size- and shape-dependent magnetic properties of $L1_0$ FePt clusters using a tight-binding approach. In their study the central plane of clusters is always Fe and they do not take into account the atomic relaxation of the clusters. Błonski and Hafner[23] undertook ab initio density-functional calculations of the magnetic anisotropy of supported nanostructures. Fernandez-Seivane and Ferrer[24] studied the correlation of the magnetic anisotropy with the geometric structure and magnetic ordering of small atomic clusters of sizes up to 7 atoms. Gruner *et al.*[18, 25] demonstrated that in cuboctahedral nanoparticles the high anisotropy of the layers increases as one moves towards the surface, and the anisotropy can be even enhanced by embedding the material in some suitable other metal (e.g., Au in the case of Pt-terminated structures). Various experimental studies, using X-ray Magnetic Circular Dichroism Spectroscopy (XMCD) or X-ray

Absorption spectra (XAS), have recently confirmed the enhancement of MCA in free or supported clusters.[26-28]

The earlier theoretical results suggest that in order to preserve the high values of the MCA, one needs to encapsulate the particles.[25] However issues coming from encapsulation such as charge transfer or structural integrity may adversely affect the MCA. In this work, we explore a possibility to tune the MCA by changing system geometry in such a way that the anisotropy mostly comes from the central part of the particle, which may help avoid the necessity of capping.

We carry out a systematic theoretical investigation of the MCA of $L1_0$ FePt clusters consisting of alternating Fe and Pt planes along the (001) direction. The clusters studied have 1(2), 2(3), 3(4) and 4(5) layers of Fe(Pt) atoms – both with Pt outer layers and with Fe outer layers – of sizes 30, 38, 71, 79, 114, 132, 140, 230, 386 and 484 atoms, respectively. We also examine the electronic structural and magnetic properties (including the orbital moments) of each atom in each of these configurations.

To calculate the structural relaxation and magnetic moments of the clusters we adopted an *ab initio* spin-polarized density functional theory (DFT) approach. To calculate the MCA we employed two methods: (i) the direct method where we take the difference in band energy for two orientations of the average magnetic moment and (ii) the torque method.[29, 30] The latter method is simpler and computationally less demanding. In this work we show its validity even for systems at the nanoscale.

We found that the MCA of layered $L1_0$ FePt clusters is enhanced over that of both bulk FePt and that of either a pure Fe or Pt cluster of comparable size, all with $L1_0$ geometry. Previous investigations attributed this enhancement is due to the hybridization 3d orbital of Fe atom with the 5d orbital of Pt atom. Our calculations indicate that this is so because this hybridization increases both spin and orbital moment of the Pt atoms. And given the large spin-orbit coupling constant of Pt it is this enhanced orbital moment of Pt that is responsible for the higher anisotropy of the system as a whole. We also found that when the central layer of the bipyramidal cluster is Pt, the cluster has higher MCA than a cluster of the same size but with Fe as the central layer, in contrast to the cuboctahedral cases,[18, 25] in which it is the surface layers that play crucial role in high MCA . This stems from the fact that when Pt atoms comprise the central layer they have more Fe atoms neighboring them, so that hybridization increases, lending them higher orbital moments than are possessed by Pt atoms in other layers. This center-of-system 'concentration' of the MCA makes it possible to preserve the anisotropy without having to resort to capping of the particles.

## II. Computational Details

To calculate the structural relaxation and spin and orbital magnetic moments we used the *ab initio* spin-polarized DFT approach implemented in the VASP code.[31] In describing the electronic exchange and correlation effects we used the spin-polarized generalized-gradient approximation (GGA) with the Pardew and Wang (PW91) functional[32] and the spin interpolation proposed by Vosko *et al.*[33] In calculating the ionic relaxation, we employed the conjugate gradient algorithm. In describing the electron-ion interaction we used the projector augmented-wave (PAW)[34] formalism. To calculate the strength of the spin-orbit coupling it is essential to take relativistic effects into account. We did so by choosing the relativistic version of PW91 as an input to the non-collinear mode framework implemented in VASP.[35, 36]

To construct the bulk FePt $L1_0$ structure we replaced every alternating layer of fcc Fe with Pt atoms in (001) planes. This ordering induces a contraction along the (001) direction of the fcc lattice, which reduces the ratio $c^*/a^*$, where $a^*$ is the nearest-neighbor distance in (001) planes (related to the primitive cell parameter $a$ as $a = a^*\sqrt{2}$), from the fcc value ($\sqrt{2}$) to 1.363. Fig. 1 shows the relation between the lattice parameter for the pseudo-cell and the parameter for the primitive unit cell.

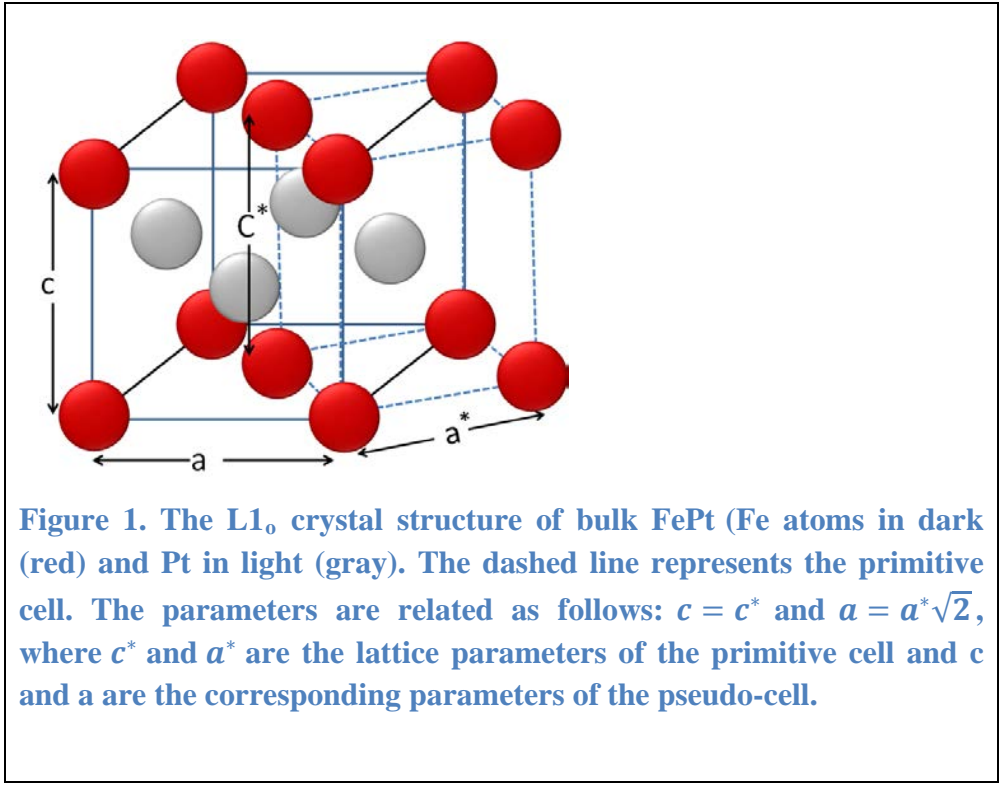

Figure 1. The $L1_o$ crystal structure of bulk FePt (Fe atoms in dark (red) and Pt in light (gray). The dashed line represents the primitive cell. The parameters are related as follows: $c = c^*$ and $a = a^*\sqrt{2}$, where $c^*$ and $a^*$ are the lattice parameters of the primitive cell and c and a are the corresponding parameters of the pseudo-cell.

Lattice distortion brings about chemical re-ordering in the unit cell. For the lattice parameter we used the experimentally-derived data for bulk powder,[37] ($a^*$=2.72Å an $c^*/a^* = 1.36$), and then relaxed the structure using the DFT approach described above.

For dimers and clusters, in relaxing the structure using the method described above, we set the vacuum space to 12 Å in all three directions, in order to prevent artificial electric field interactions between the images, and used only one (Gamma) point in the Brillouin zone. To obtain the relaxed geometry of a cluster of given size and shape, we first obtained the relaxed lattice parameter for bulk $L1_0$ FePt, then cut the bulk to the size and layered shape of the cluster in question to get its initial configuration, and finally relaxed that configuration. In calculating the MCA of $L1_0$ FePt nanoparticles we used both the direct method[23] (i.e., including spin-orbit coupling self-consistently), and the torque method.[29,30]

In the direct method, once the structural relaxation for a given cluster is completed, we calculate the MCA by comparing the band energy between two magnetic orientations. The band energy of the system is calculated by using spin-polarized DFT, now taking SOC into account.

$$MCA = E(\uparrow) - E(\rightarrow) = \sum_{occ'} \varepsilon_i(\uparrow) - \sum_{occ''} \varepsilon_i(\rightarrow) \qquad (1)$$

where $\varepsilon_i$ is the band energy of the $i^{th}$ state and the arrow in the parentheses denote the direction of magnetization. (For the surface of a film, these are usually the directions perpendicular and parallel to the film and for nanoparticles they are the directions along the z-axis and parallel to xy-plane, respectively). We aligned the magnetization towards two mutually perpendicular directions setting spin arrangement in the system to be non-collinear. The sums in equation (1) are usually large numbers (on the order of hundreds of eV), whereas the MCA is on the order of few meV. Since MCA is a small number coming from the difference of two large numbers, one needs to take great care, in determining the Kohn-Sham energy of the system, in selecting the convergence criterion for the calculation. In fact, for accurate integration, it is necessary to use a very fine k-point mesh in reciprocal space. And the fact that a very accurate convergence of energy is also required in the self-consistent cycle makes the calculations rather expensive. This method can give quite divergent results if one does not use sufficient number of integration points: for example Gay and Richter[38] predict the easy axis of monolayer Fe to be perpendicular to the plane and found the anisotropic energy to be -0.4 meV/atom, whereas Karas *et at.*[39] report the easy axis of the same system to be along the plane of the layer, and the value of anisotropy to be 3.4 meV/atom. Since we use a large supercell in the calculations, we do not need to include large number of k points, though we did carry out a set of calculations for 3x3x3 k-point mesh as a test for our 38-atom clusters, and the results are almost same as that obtained with the Γ point. Therefore, in other calculations we confined ourselves to using simply the Γ point.

An alternative way of calculating the MCA is the torque method[29, 30] which is suitable for systems with uniaxial symmetry. The advantages of this method are that it does not require calculations of the total energy of the system with very high accuracy and that it is much faster because it does not require self-consistent calculations with SOC for two different directions of the magnetization. For our calculations using the torque method we employed the VASP post-processing package developed by Jun Hu et al.,[40] based on the augmented-wave projection of the SOC operator[41]. We also calculated the MCA to see whether the results differed considerably from those obtained from the direct method. (As Figure 7 indicates, they generally did not.)

## III. Results and Discussion

### Magnetic anisotropy of the bulk system

The lattice parameters of the relaxed structure of bulk $L1_0$ FePt do not change much from the experimental ones ($a^*$=2.72 Å and $c^*/a^* = 1.36$): the in-plane parameter $a^*$=2.74 Å, and the ratio value $c^*/a^* = 1.37$. We have obtained the following values for the magnetization: 2.85 $\mu_B$ for the Fe atoms and 0.36 $\mu_B$ for the Pt atoms, both in good agreement with experiment (Fe: 2.90 $\mu_B$, Pt: 0.34 $\mu_B$).[37] It is worth mentioning that though bulk Pt is nonmagnetic, in the FePt alloy the Pt atoms possess a magnetic moment. The magnetic moment of Fe atoms increases substantially in the alloy from their values in bulk Fe. The enhancement of the Fe magnetic moment is a consequence of the hybridization between the states of the 3d orbitals of the Fe atoms and not only the 5d but also s- and p-orbitals of their neighboring Pt atoms. As can be seen in Fig. 2, the hybridization causes a broadening of the d-bands for the majority spins, while its effect on the shift of the minority spins, though smaller, that shift moves the band across the Fermi level into the area unoccupied by any electron. Thus, it is natural to suppose that the finite magnetic moment (0.36 $\mu_B$) of Pt atoms in FePt clusters comes mostly from the hybridization of the minority spin bands.

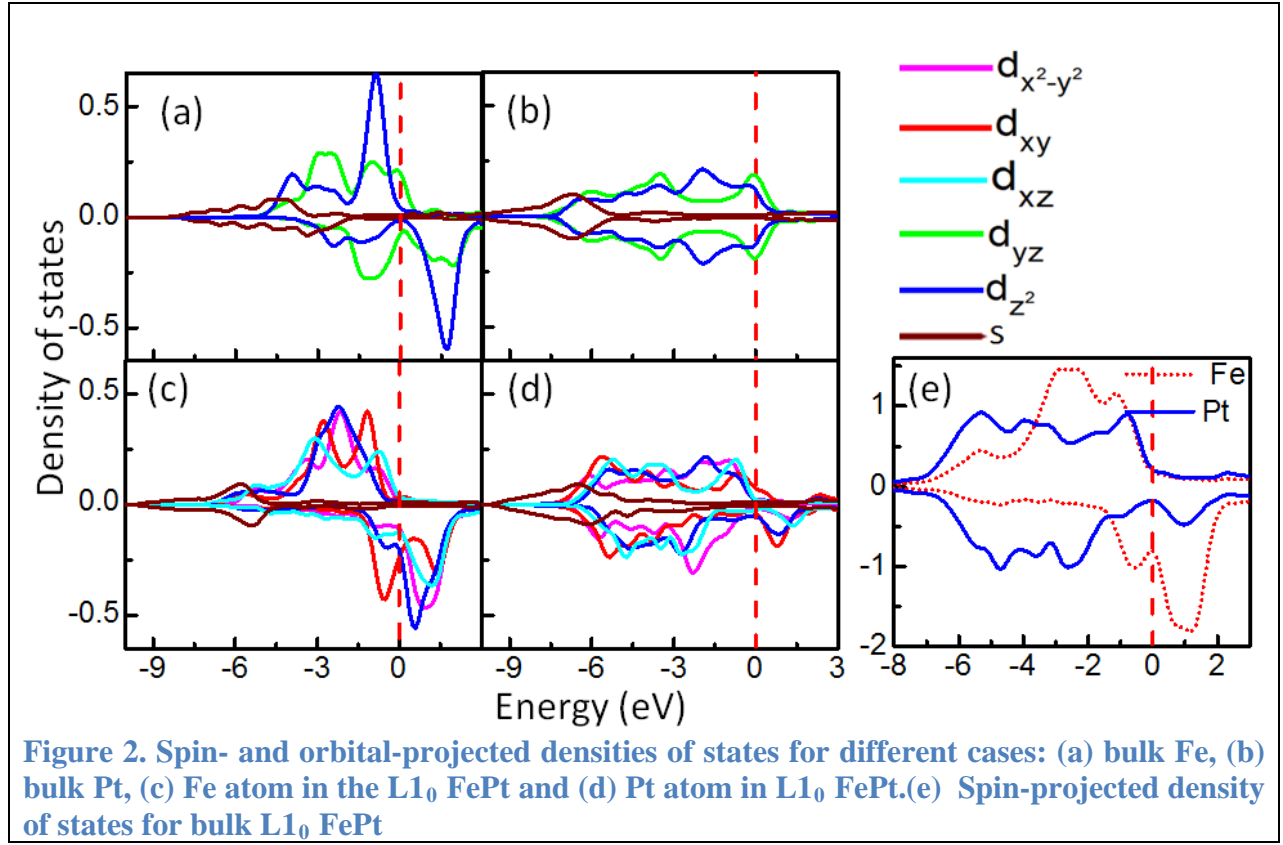

Figure 2. Spin- and orbital-projected densities of states for different cases: (a) bulk Fe, (b) bulk Pt, (c) Fe atom in the L1$_0$ FePt and (d) Pt atom in L1$_0$ FePt.(e) Spin-projected density of states for bulk L1$_0$ FePt

The density of states (DOS) of bulk L1$_0$ FePt is plotted in Fig. 2, where we also present the projected DOS for the Fe and Pt atoms, which is similar to the one reported by C. Barreteau *et al.*[22] Our calculations show MCA value for the bulk to be 2.22 meV per FePt pair, which is also in agreement with other studies,[2] encouraging confidence that our results in the nanocase are reliable as well. We attribute the large MCA to the large SO coupling of Pt atoms and to the increase in orbital moment owing to the strong hybridization of their 5d orbitals with the 3d orbitals of Fe. It is this hybridization that breaks the symmetry of free energy in the two perpendicular directions, resulting in a surplus of free energy in the direction of magnetization.[19] Interestingly, Lyubina *et al.* found that the anisotropy of the disordered phase of FePt is an order of magnitude less than that of the L1$_0$ phase.[11]

## Magnetic anisotropy of the dimer

To gain more insight into the nature of MCA in the nanoparticles that are the chief object of our study, we also considered the case of pure Fe, pure Pt, and FePt dimers. For Fe$_2$ the bond length is found to be 1.98 Å and that for Pt$_2$ the bond length is 2.37 Å which are in good agreement with the bond length reported in the earlier work by Blonski et al.[23] and the references therein. The experimental data is available only for the Fe dimer (bond length is 1.87 Å)[42] which is also in good agreement with our calculations. For the FePt dimer we obtained a dimer bondlength of 2.18 Å when leaving spin-orbit coupling out of account, and 2.17 Å when taking SOC into account. We thus infer that SOC does not play an important role in determining the geometry of this system. This result is also in good agreement with the results of Ref.[43] Fe$_2$ turns out to exhibit very small anisotropy owing to the relatively small SOC of the Fe atom. The SOC of Fe atom is 81.6 meV and of the Pt atom is nearly 7 times greater 544 meV[44]. Contrary to the other two dimers, in Pt$_2$ the spin moment and orbital moment significantly differ in the directions along the dimer and perpendicular to it. Therefore, it is the Pt atom's high SOC and relatively large *orbital* moment in Pt$_2$ that together account for the fact that the MCA of that dimer greatly exceeds those of the other two. The MCA of the mixed FePt system has a value - |10.37| meV - between those of the two monometallic cases - |0.07| and |55.94| .meV, respectively. Table 1 and Figure 3 help us understand why this is the case. In the FePt dimer the total magnetic moment of the Fe atom is 3.22 $\mu_B$ and that of the Pt atom - 0.58 $\mu_B$. The large magnetic moment of Fe in this dimer can be attributed to the charge transfer from the 3d orbital of Fe to the 5d orbital of the Pt atom, which creates a polarization and an extra "hole charge" on the Fe atom.[43] The MCA for FePt dimer is 10.37 meV (with the easy axis perpendicular to the dimer axis). This value is higher than the value for Fe$_2$, because the values of both the spin and orbital momenta are larger in the FePt dimer. Our results for the spin and orbital momenta and the MCA energy along with corresponding available numerical results obtained by other methods are compiled in Table 1.

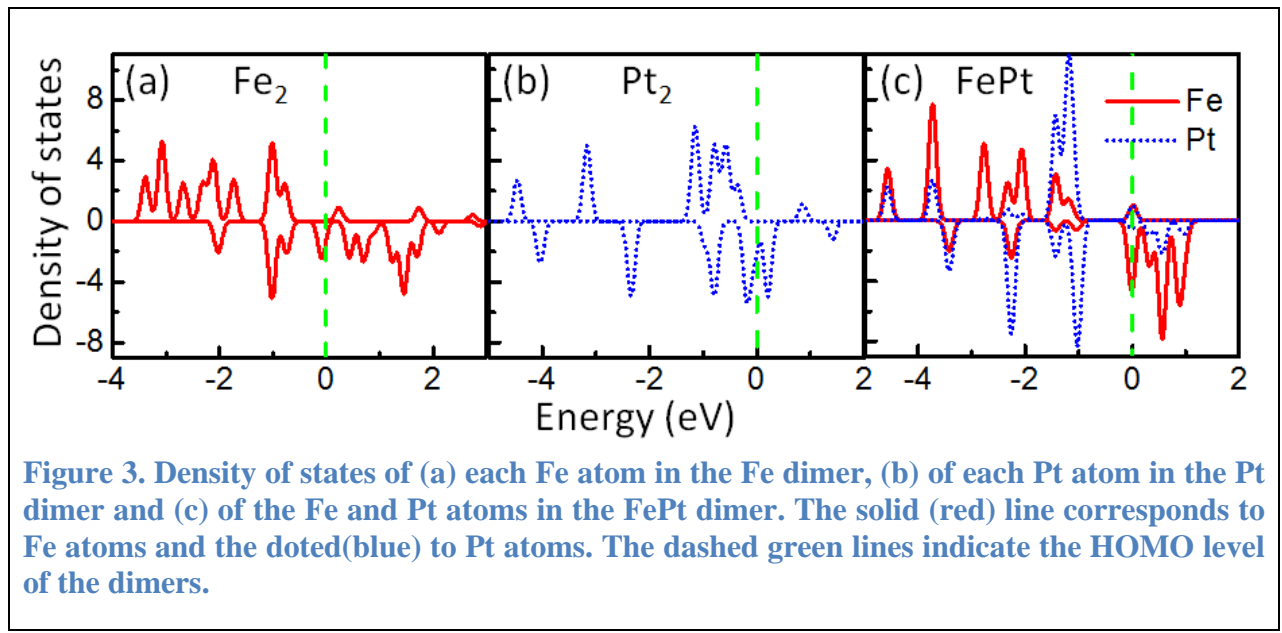

Figure 3. Density of states of (a) each Fe atom in the Fe dimer, (b) of each Pt atom in the Pt dimer and (c) of the Fe and Pt atoms in the FePt dimer. The solid (red) line corresponds to Fe atoms and the doted(blue) to Pt atoms. The dashed green lines indicate the HOMO level of the dimers.

In all three cases, the easy axis of magnetization coincides with the direction of highest orbital momentum, in agreement with Bruno[45] (See Table 1). The negative MCA for the case of the FePt dimer means that the easy axis of magnetization is perpendicular to the dimer axis in this case.

Table 1: Magnetic moments and MCA energy (in meV) of the Fe, Pt and FePt dimers. LDA stands for the local density approximation in DFT.

| Dimer | | Spin Moment ($\mu_B$) | | Orbital Moment ($\mu_B$) | | MCA |
|---|---|---|---|---|---|---|
| | | (100) direction | (001) direction | (100) direction | (001) direction | $E_{100}-E_{001}$ |
| $Fe_2$ | This work | 5.99 | 5.99 | 0.32 | 0.16 | 0.07 |
| | GGA[23] | 5.84 | 5.84 | 0.32 | 0.16 | 0.3 |
| | LDA[46] | 6.00 | 6.00 | 1.89 | 0.89 | 32 |
| | GGA[43] | 6.00 | 6.00 | 0.25 | 0.10 | 0.5 |
| $Pt_2$ | This work | 1.89 | 1.38 | 2.74 | 0.80 | 55.94 |
| | GGA[23] | 1.88 | 1.34 | 2.74 | 0.80 | 46.3 |
| | LDA[47] | 1.90 | 1.65 | 2.40 | 1.20 | 220.0 |
| | GGA[43] | 1.80 | 1.70 | 2.40 | 0.80 | 75.00 |
| FePt | This work | 4.16 | 4.26 | 0.36 | 0.41 | -10.37 |
| | GGA[43] | 4.30 | 4.30 | 0.2 | 0.40 | |

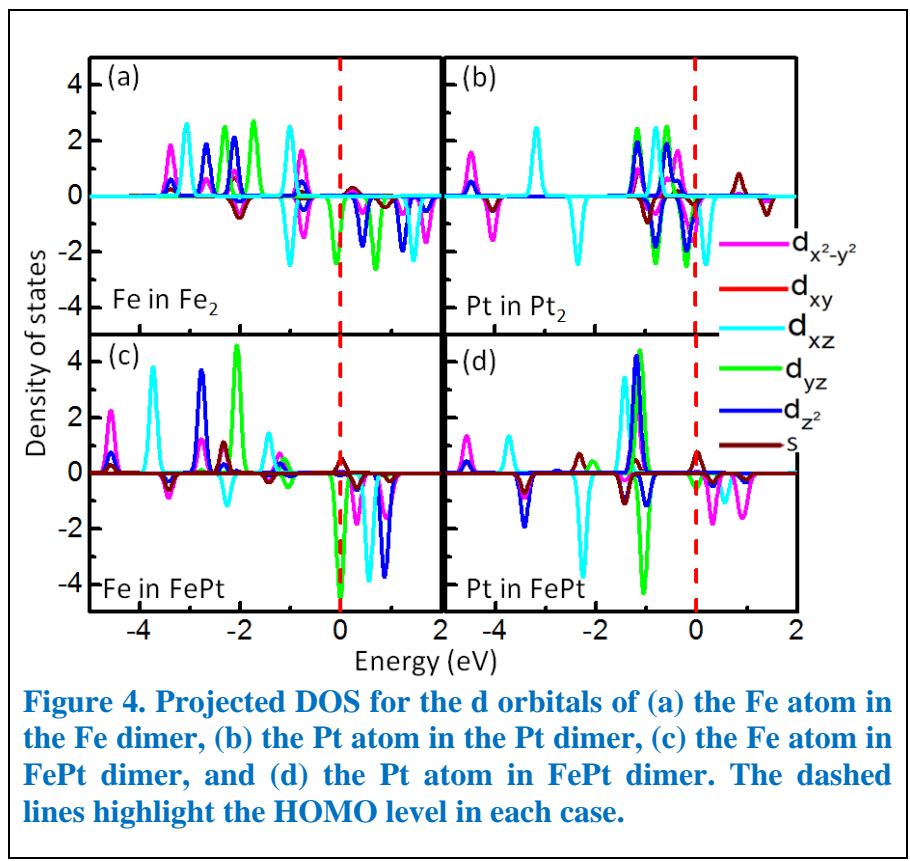

**Figure 4. Projected DOS for the d orbitals of (a) the Fe atom in the Fe dimer, (b) the Pt atom in the Pt dimer, (c) the Fe atom in FePt dimer, and (d) the Pt atom in FePt dimer. The dashed lines highlight the HOMO level in each case.**

The projected DOS for the $d$ orbitals of the Fe and Pt atoms in the dimers are shown in Fig.4. As follows from this figure, the majority spin orbitals have zero occupancy at the HOMO level, but the minority spins are present there, indicating that the change of the orbital occupancy upon hybridization is defined by the charge transfer for the spin-down electrons only. We thus conclude that the anisotropy and the magnetization of the dimers are generated by the minority spins as well.

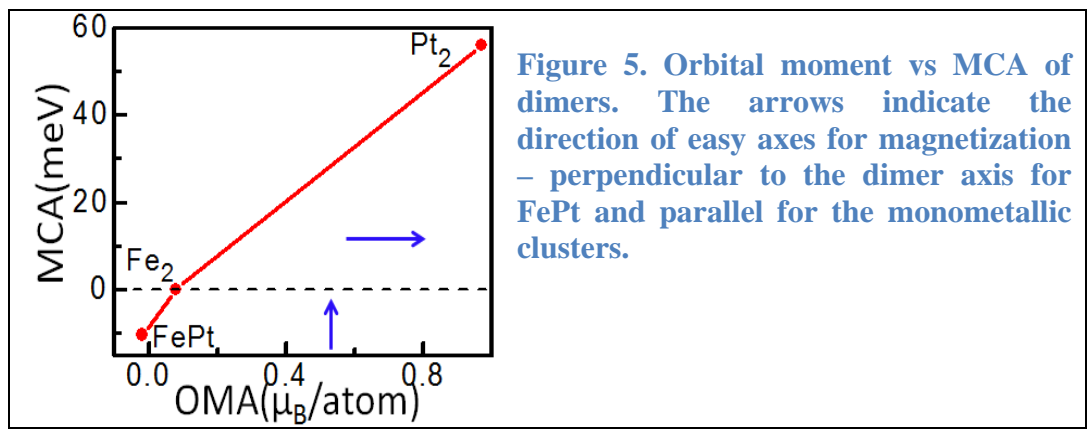

Figure 5. Orbital moment vs MCA of dimers. The arrows indicate the direction of easy axes for magnetization – perpendicular to the dimer axis for FePt and parallel for the monometallic clusters.

To summarize, the MCA is proportional to the change of orbital momentum in two different directions; the comparatively small value of MCA for $Fe_2$ is due to its small SOC constant; larger value of MCA for $Pt_2$ is due both to its large SOC constant and to its relatively large orbital moment; while the intermediate value of MCA in the FePt dimer is due to mutual tempering of each atom of the other's SOC and to the fact that its orbital moment falls between those of $Fe_2$ and of $Pt_2$.

## Magnetic anisotropy of $L1_0$ FePt nanoclusters

The high MCA of bulk $L1_0$ FePt alloy with equiatomic composition has stimulated researchers to inquire into the properties of its small particles. To be sure, there are some problems in maintaining the bulk geometry for the nanoparticles. For example, Muller *et al*. predicted theoretically that the $L1_0$ phase is not thermodynamically stable[48], and another study, by Jarvi *et al*.[49] showed that the structure may alter, disrupting the original atomic order. Moreover, several experimental studies support the existence of a minimum size limit below which the $L1_0$ order can no longer be achieved,[50, 51] and other studies have shown the migration of Pt atoms towards the surface in smaller particles.[52, 53] However, for more definitive answers concerning feasibility of exploiting FePt nanoparticles in the applications mentioned in the Introduction, detailed and systematic studies are in order.

Those undertaken so far are suggestive. In particular, it is known that the magnetic properties of nanoparticles depend on the size, shape and the way of synthesizing.[11] Gas-phase particles can in fact exhibit *lower* MCA than perfectly-ordered bulk $L1_0$ alloys, because (i) their internal structure may not be perfect $L1_0$, and individual particles can become multiply-intertwined, (ii) the Pt atoms may tend to migrate towards the surface,[43,44] or (iii) an inhomogeneous alloying may be present from the beginning, as indicated by the EXAFS measurements by Antoniak.[54] The enhancement of surface-to-volume ratio (and hence the size) of a nanoparticle plays a significant role in its MCA. For example, the crystal symmetry-dependent quenching of the orbital magnetic moments disappears for all surface atoms of nanoparticles, thereby enhancing their orbital moments relative to those of bulk or core atoms.[55] (Antioniak's XMCD measurements have confirmed this for Fe surface atoms in $L1_0$ FePt nanoparticles.[56])

The cluster structures we studied – constructed as described in the section on Computational Details – are shown in the Fig. 6.

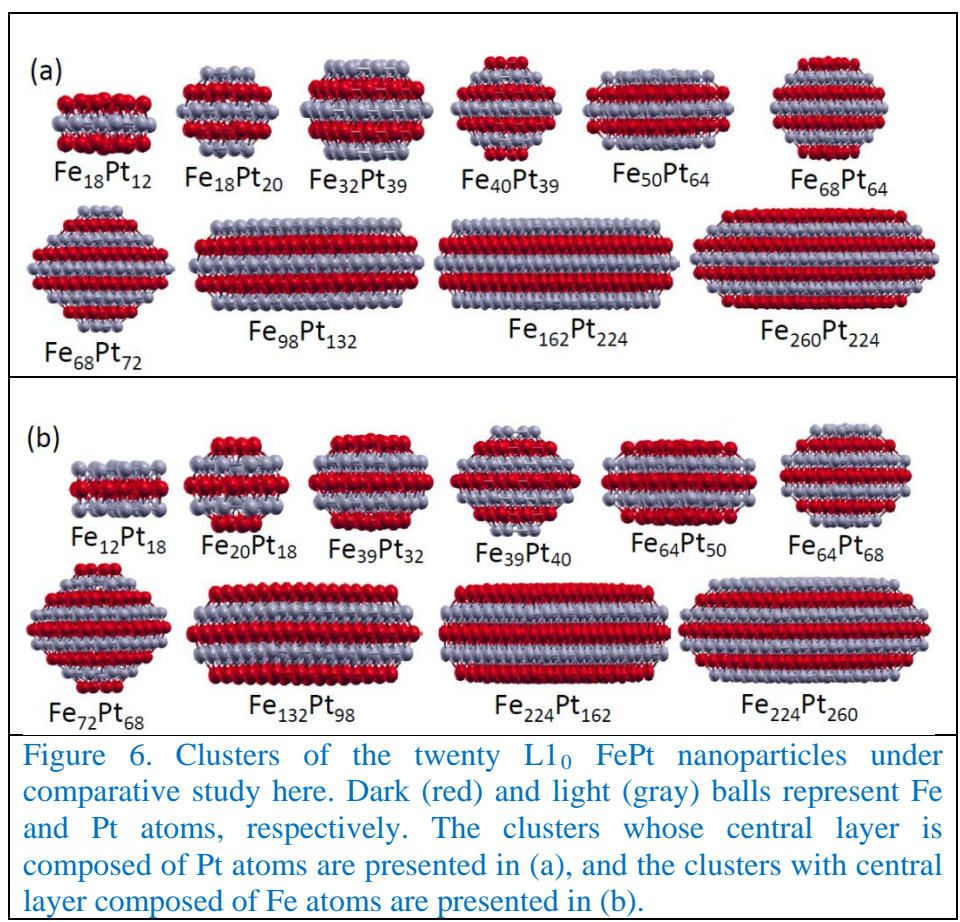

Figure 6. Clusters of the twenty $L1_0$ FePt nanoparticles under comparative study here. Dark (red) and light (gray) balls represent Fe and Pt atoms, respectively. The clusters whose central layer is composed of Pt atoms are presented in (a), and the clusters with central layer composed of Fe atoms are presented in (b).

The values of MCA for the above clusters obtained with both methods are presented in Fig. 7.

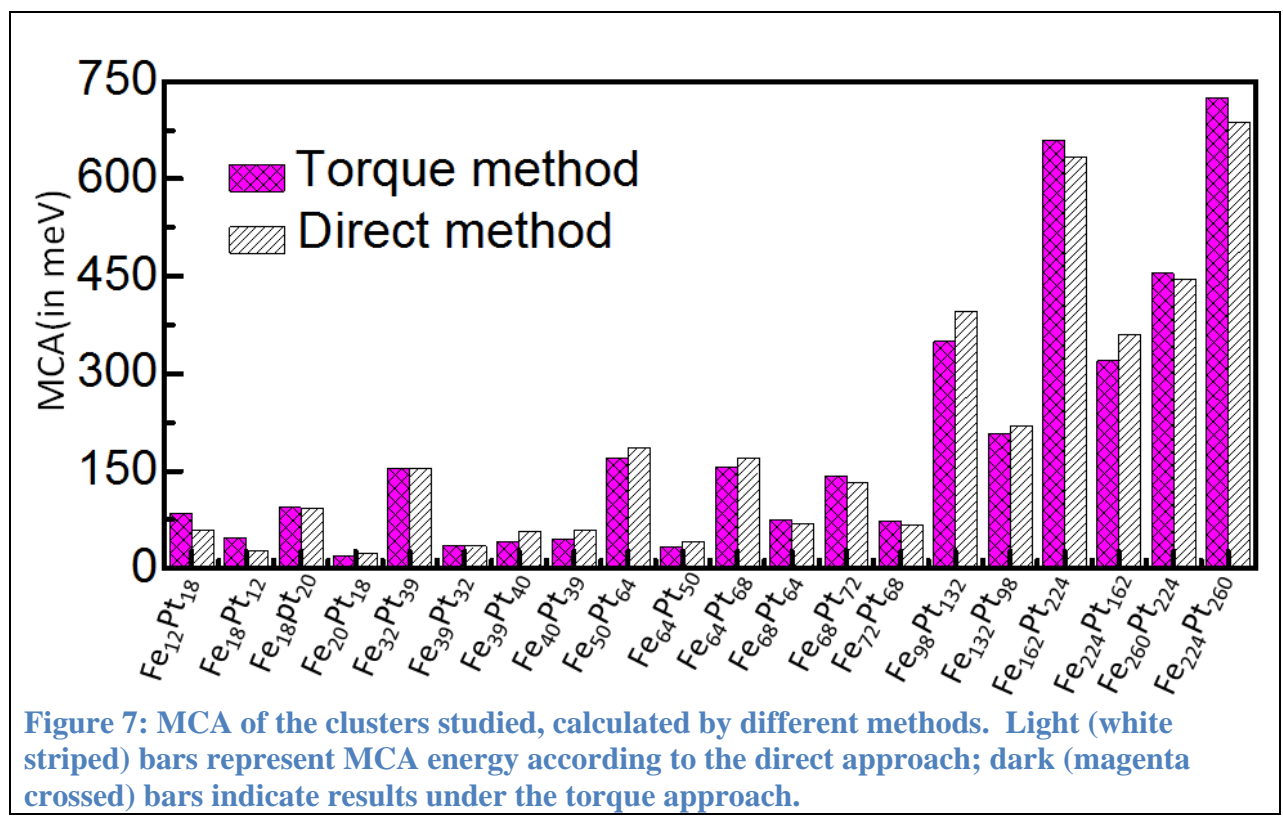

Figure 7: MCA of the clusters studied, calculated by different methods. Light (white striped) bars represent MCA energy according to the direct approach; dark (magenta crossed) bars indicate results under the torque approach.

The results for MCA calculated with the direct and the torque methods generally agree well. This is encouraging, particularly for nanostructures, because the direct method in this case is much more computationally demanding. One definite conclusion supported by the calculations is that those nanoparticles with a larger number of Pt atoms have larger MCA than their counterparts with a larger number of Fe atoms. We also find that the atoms on the outside of the clusters have higher magnetic moments than do atoms inside. One example of such a situation is shown in Fig.8 for $Fe_{72}Pt_{68}$ and $Fe_{68}Pt_{72}$ clusters. The line through which we calculate the magnetic moment is presented by the arrow in Fig. 8(a). The line (red) with solid circles and the line (blue) with open circles in Fig. 8(b) represent magnetization for Fe (middle layer with Fe atoms, $Fe_{72}Pt_{68}$) and Pt (middle layer with Pt atoms, $Fe_{68}Pt_{72}$) atoms, respectively. It is clear from the figure that the outer atoms in the cluster have larger magnetization than the inner atoms. As we shall see, this comes from the fact that exterior atoms have fewer neighbors than do those inside the cluster.

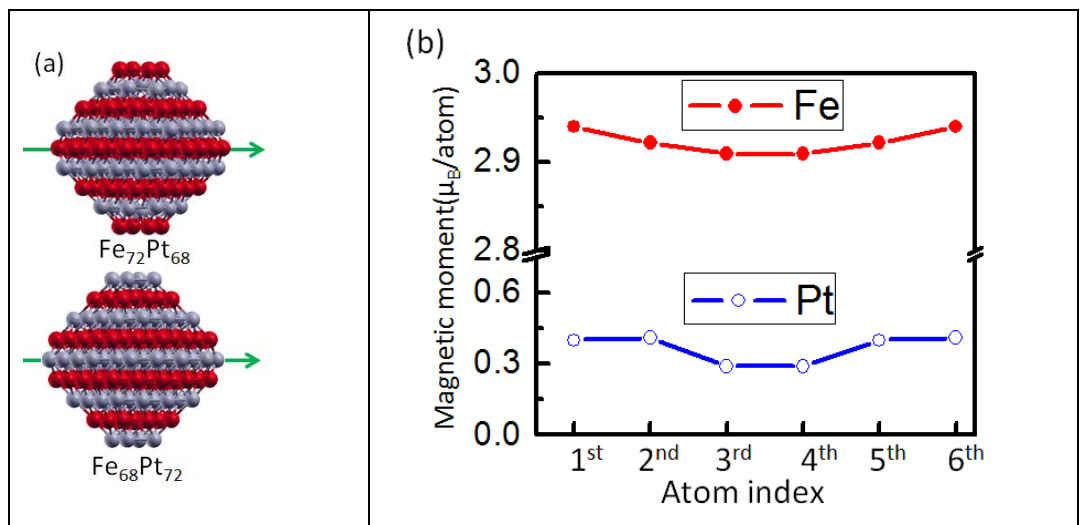

**Figure 8:** Magnetic moment of atoms at different positions in central layer of the cluster. The arrows in (a) indicate the row in which we picked atoms for comparing their magnetization. (b) The magnetic moment of atoms at different positions. For this pair of clusters, there are six atoms along the line passing through the cluster's center; the 3$^{rd}$ and 4$^{th}$ atoms are at the center of the cluster.

Pt atoms in the FePt clusters exhibit magnetization, in contrast to atoms in bulk Pt (where they exhibit virtually none). Once again, the inside Pt atoms have smaller magnetization than do the Pt atoms on the outside. The contrast between Fe atoms in our FePt clusters and Fe atoms in bulk Fe is different but parallel: Atoms in bulk Fe do show magnetization, but Fe atoms in FePt clusters exhibit even higher magnetization. Here, too, the inside atoms exhibit less magnetization than those on the outside of the cluster. All four contrasts – between magnetization of atoms pure bulk and in composite clusters, and between that of interior and exterior atoms within clusters of the same size – are due to the orbital hybridization of Pt with Fe atoms. The lower number of neighbors for surface atoms explains the narrowing of the 3$d$ orbital bands of the Fe and the 5$d$ bands of Pt that is in turn responsible for enhancing the magnetization of the surface atoms.

The projected DOS for the atoms at different positions in Fe$_{20}$Pt$_{18}$ and Fe$_{18}$Pt$_{20}$ clusters is presented in Fig. 9.

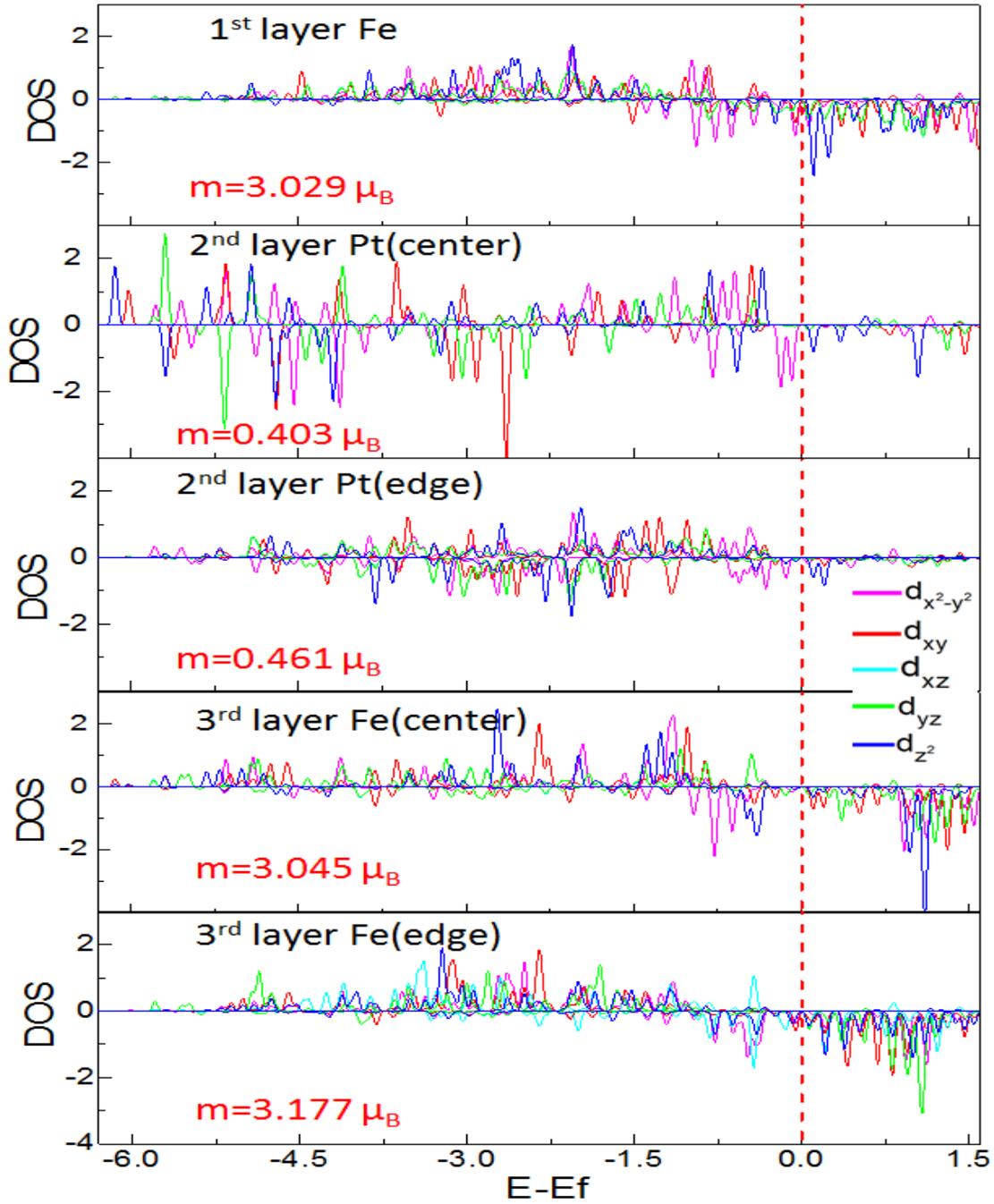

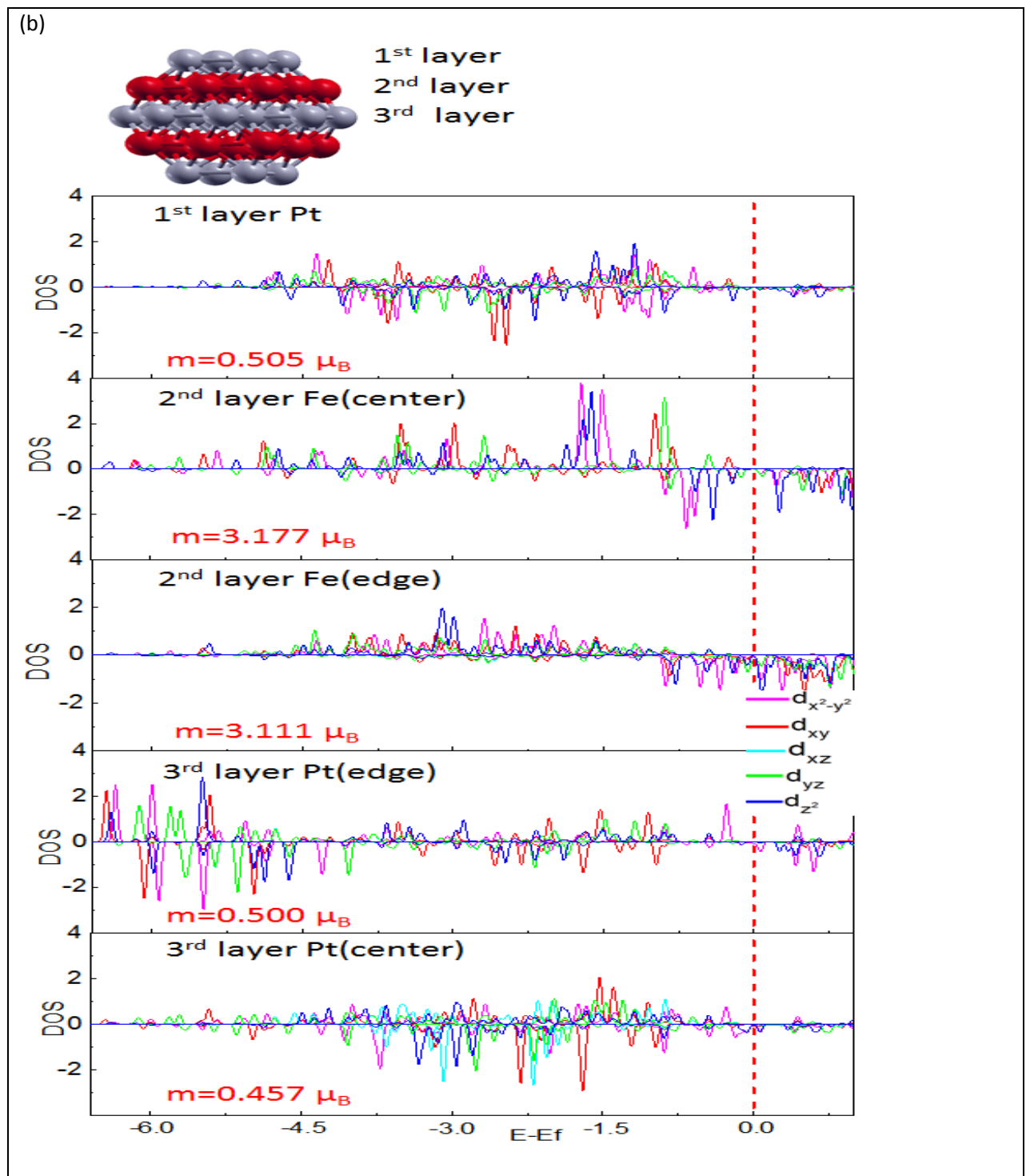

**Figure 9.** Projected density of states for the atoms at different positions, (a) in $Fe_{20}Pt_{18}$ and (b) in $Fe_{18}Pt_{20}$ cluster. The DOS is more localized for an atom on the surface of the cluster than for one atom inside. For all cases the majority spin band is completely filled and magnetization is due to the contribution from the minority band only. The value of m in each graph is the magnetic moment of that particular atom.

As this figure makes clear, there is a significant difference in the PDOS for the Pt atoms within the interior and at the periphery of the $Fe_{18}Pt_{20}$ cluster. Remarkably, the main contribution to this difference comes from the less filled spin-down $d_{x^2-y^2}$ orbital in the case of the inside Pt atoms. This leads to a significant difference in the orbital momenta of the inside and outside atoms and hence of the high MCA. On the other hand, in the case of "flipped" cluster $Fe_{20}Pt_{18}$ one gets less difference in the orbital occupancies for the inside and outside atoms for both cases of Fe and Pt. Thus, it is not surprising that we find that bi-pyramidal structures with large central Pt layer has significantly higher MCA per atom compared to the cuboctahedral ones.[18, 25] For example, we get an MCA per atom value 1.14 meV for the $Fe_{64}Pt_{68}$ versus 0.86 meV reported in the papers above for the 147 atom FePt cluster reported in Ref [25]. This difference can be traced to the larger percentage of Pt atoms in the center for the bi-pyramidal case compared to the cuboctahedral structure.

**Table 2: The magnetic moments, orbital moments, and MCA of clusters with magnetization along two different directions.**

|  | Magnetic Moment ($\mu_B$) | | Orbital Moment ($\mu_B$) | | MCA (meV) |
| --- | --- | --- | --- | --- | --- |
|  | $m_x$ | $m_z$ | $l_x$ | $l_z$ | MCA=$(E_x-E_z)$ |
| $Fe_{12}Pt_{18}$ | 46.74 | 46.26 | 2.283 | 2.037 | 84.60 |
| $Fe_{18}Pt_{12}$ | 62.33 | 62.03 | 2.14 | 2.54 | 47.60 |
| $Fe_{18}pt_{20}$ | 68.16 | 68.06 | 3.31 | 2.97 | 94.37 |
| $Fe_{20}Pt_{18}$ | 72.76 | 72.75 | 3.30 | 3.09 | 20.01 |
| $Fe_{32}Pt_{39}$ | 117.18 | 115.19 | 5.86 | 4.46 | 154.28 |
| $Fe_{39}Pt_{32}$ | 134.18 | 134.14 | 5.39 | 5.22 | 34.15 |
| $Fe_{39}Pt_{40}$ | 133.57 | 132.46 | 5.74 | 5.17 | 81.05 |
| $Fe_{40}Pt_{39}$ | 137.26 | 136.53 | 6.13 | 5.63 | 71.82 |
| $Fe_{50}Pt_{64}$ | 175.97 | 175.67 | 7.95 | 6.91 | 170.11 |
| $Fe_{64}Pt_{50}$ | 213.24 | 213.44 | 8.52 | 7.91 | 33.33 |
| $Fe_{64}Pt_{68}$ | 214.84 | 214.76 | 9.64 | 8.24 | 155.90 |
| $Fe_{68}Pt_{64}$ | 232.88 | 232.64 | 10.08 | 9.11 | 75.06 |
| $Fe_{68}Pt_{72}$ | 233.90 | 231.71 | 10.59 | 10.06 | 142.80 |
| $Fe_{72}Pt_{68}$ | 242.69 | 242.95 | 10.15 | 10.51 | 73.65 |
| $Fe_{98}Pt_{132}$ | 350.02 | 350.78 | 16.90 | 14.57 | 349.38 |
| $Fe_{132}Pt_{98}$ | 452.21 | 451.98 | 15.86 | 13.42 | 340.99 |
| $Fe_{162}Pt_{224}$ | 583.97 | 583.97 | 28.11 | 23.63 | 659.77 |
| $Fe_{224}Pt_{162}$ | 722.66 | 722.07 | 27.47 | 26.32 | 320.27 |
| $Fe_{260}Pt_{224}$ | 870.00 | 870.21 | 36.10 | 33.38 | 454.44 |
| $Fe_{224}Pt_{260}$ | 773.98 | 773.94 | 34.581 | 30.38 | 725.53 |

As Table 2 reveals, the total magnetic moment does not significantly change as the direction of magnetization shifts from (001) to (100). The orbital moment changes in such a way that the easy axis of magnetization is always along the direction of the lower orbital moments. This finding contradicts Bruno's prediction,[45] according to which the easy axis of magnetization always coincides with the direction of the highest orbital moments. To be sure, for dimers (Table 1) the direction of easy magnetization does follow the direction of highest orbital moment of the system, but for larger clusters it behaves in the opposite fashion. One would expect the easy axis of magnetization of still larger clusters to exhibit the same sort of alignment of direction of easy axis of magnetization with the direction highest orbital moment, as in those clusters under study here..

The magnitude of difference in orbital moment between the directions of magnetization $\Delta l = l_x - l_z$ *per atom* also plays a key role in determining the overall MCA of clusters. As Table 2 shows, for the majority of clusters anisotropy increases with increase in the difference between orbital moment per number of atoms. And, between clusters of the same size, anisotropy is higher in the cluster those whose predominant constituent element has the higher SOC energy. In sum: for majority of clusters we studied, the contribution to MCA comes mostly from increase in the orbital moment of the system along with increase in SOC energy, though in general case the dependence of the MCA on the size and shape of the clusters is highly nontrivial and requires further detailed studies.

We also studied the MCA of different layers within the clusters (Fig. 10).

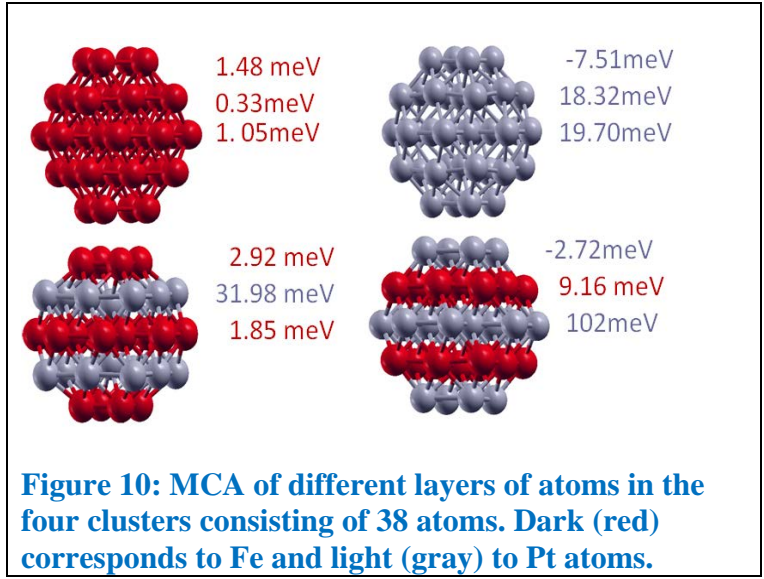

**Figure 10: MCA of different layers of atoms in the four clusters consisting of 38 atoms. Dark (red) corresponds to Fe and light (gray) to Pt atoms.**

As Fig. 10 indicates, the central plane of a cluster of the same size has much higher anisotropy when it consists of Pt than when it consists of Fe atoms. Anisotropy is also higher when the central Pt layer adjoins a layer of Fe than when its neighboring layer is Pt. This turns out to be the case for all of the clusters we studied which can be explained as follows. When a Pt atom hybridizes with neighboring Fe atoms in FePt clusters, the orbital moment of Pt atoms increases. In the example above, the Pt atom in the $Fe_{20}Pt_{18}$ cluster shows an orbital moment of 0.11 $\mu_B$ but in $Fe_{18}Pt_{20}$ its orbital moments increases to 0.145 $\mu_B$. This increase in the orbital moments, along with the greater SOC energy of Pt, thus increases the MCA of the central layer. These values of the orbital moment for the Pt atoms are much higher than that of Pt in $Pt_{38}$ (0.02 $\mu_B$), Fe in $Fe_{38}$ (0.06 $\mu_B$), Fe in $Fe_{20}Pt_{18}$ (0.06 $\mu_B$), and Fe in $Fe_{18}Pt_{20}$ (0.04 $\mu_B$). The same pattern holds for all sizes of clusters we study here.

The difference in anisotropy for different layers of Fe/Pt atoms can be explained in simple terms of orbital occupancies as we show below from perturbation theory. This is appropriate because the portion of the $d$ bandwidth contributed by the SOC component of the Hamiltonian is much smaller than the $d$ bandwidth as a whole.[48] We see from Fig. 9 that, since all the majority spin states are completely occupied, all the empty states belong to the minority spins only. Since there are no empty states available for occupation by the spin-up electrons, there are only two types of SO interactions in the systems: the coupling between occupied and unoccupied spin-down states and the coupling between states occupied by spin-up electrons with states unoccupied by spin-down electrons. The MCA energy can be calculated by using the following formula:[48]

$$E_X - E_Z \sim \xi^2 \sum_{o,u} \frac{|\langle o|\hat{L}_z|u\rangle|^2 - |\langle o|\hat{L}_x|u\rangle|^2}{\varepsilon_u - \varepsilon_o} \qquad (2)$$

where $\xi$ is the spin-orbit coupling constant, $\langle O|$ and $\langle u|$ are the occupied and unoccupied minority spin states, respectively, and $\hat{L}_z$ and $\hat{L}_x$ are the z and x components of the angular momentum operators. We have calculated the orbital occupancy of both empty and filled states of both spin arrangements and find that for the Fe atom in the 1$^{st}$ layer of Fe$_{20}$Pt$_{18}$ $E_x - E_z = 0.24\xi^2$, for the atoms in the 3$^{rd}$ layer (exterior) of Fe$_{20}$Pt$_{18}$ this value is $E_x - E_z = 0.17\xi^2$ and for the atoms in the 3$^{rd}$ layer(interior) of Fe$_{20}$Pt$_{18}$ this value is $E_x - E_z = 0.08\xi^2$.

Thus, our perturbation theory estimation from the last paragraph gives a MCA per atom for the top Fe layer that is 2.5 times larger than the corresponding value for the third Fe layer. This result is in a good agreement with the DFT ratio for the MCA's, ~3, for the corresponding layers (see Fig.10), suggesting the ability of DFT to describe correctly the MCA in these clusters in terms of explicit orbital occupancies. The discrepancy in results for the MCA may come from the simplicity of the estimation used. For greater accuracy of the perturbation theory calculations, one would need to take into account the differences in hybridization undergone by each individual atom. An example of the orbital occupancies for differently-situated atoms is given in the Table 3. Indeed, as it follows from this Table, the occupancies of individual d-orbitals are much less than one, which suggests strong hybridization of these orbitals.

**Table 3: Occupancy of d-orbitals of differently situated atoms in the 1$^{st}$ and 3$^{rd}$ Fe layers in the Fe$_{18}$Pt$_{20}$ cluster.**

| Projected d-orbitals | 1$^{st}$-layer atom (all exterior) | 3$^{rd}$-layer atom (exterior) | 3$^{rd}$-layer atom (interior) |
|---|---|---|---|
| $d_{x^2-y^2}$ | 0.51 | 0.45 | 0.45 |
| $d_{xy}$ | 0.29 | 0.17 | 0.25 |
| $d_{xz}, d_{yz}$ | 0.30 | 0.10 | 0.16, 0.13 |
| $d_{z^2}$ | 0.18 | 0.28 | 0.28 |

## IV. Thermal Effects

The calculations above are all appropriate for zero temperature. We now deal with two thermal effects. The first topic deals with the potential application for FePt nanoparticles in the magnetic storage of data. It is well known that many nano-sized particles are superparamagnetic because their total anisotropy energy is on the order of the thermal energy. A standard measure for the magnetic stability time of a small magnetic particle is given by an Arrhenius law[57] involving the probability of climbing over an energy barrier. For magnetic storage applications, the ratio of the anisotropy energy to thermal energy is quite large. For T = 25 years, one finds

$$\frac{KV}{k_B T} = ln\left(T/T_0\right) \geq 41. \tag{3}$$

Here K is the anisotropy per unit volume and V is the volume. We have used $T_0 = 10^{-9} sec$ as the value for a typical attempt time for the system to climb the anisotropy barrier. With our values for the largest anisotropy (KV is about 725 meV for the Fe$_{224}$Pt$_{260}$ cluster), we get

$$\frac{KV}{k_BT} = 28 \tag{4}$$

Clearly, the total anisotropy energy is still not large enough compared to the thermal energy. The required anisotropy energy for the particle is about 1060 meV to satisfy the condition in Eq (3). This means that one still needs larger particles if these elements are to work for magnetic storage. We can use our previous data to project values of MCA for larger particles. It follows from Fig. 7 that the MCA of clusters with the same number of layers scales nearly linearly with the number of Pt atoms in the cluster. As an example we plot the MCA as a function of the number of Pt atoms for 5 layer clusters with Pt as central layer in Fig. 11.

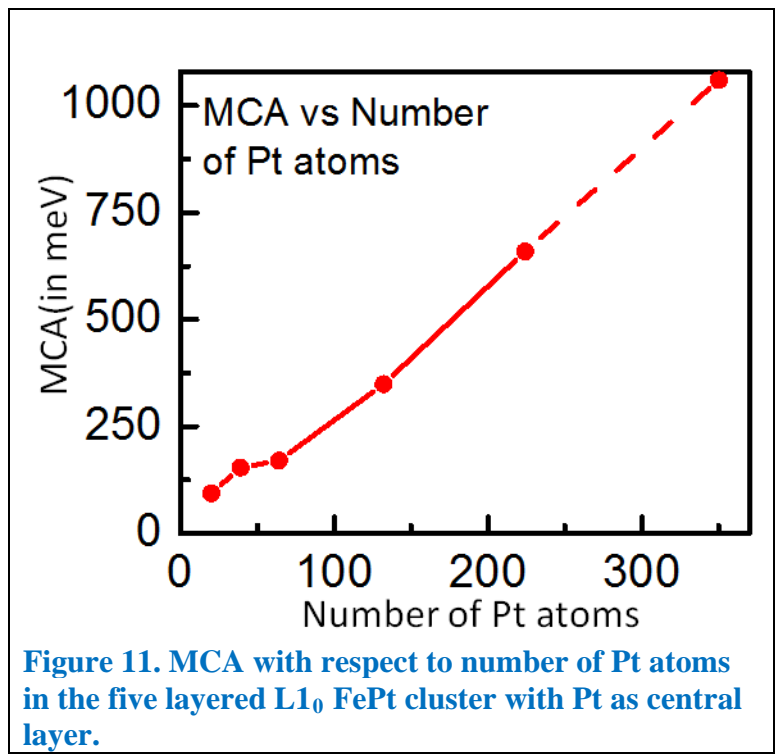

**Figure 11. MCA with respect to number of Pt atoms in the five layered L1$_0$ FePt cluster with Pt as central layer.**

From Fig. 11 we can predict (for this structure) that to get an MCA of 1060 meV per cluster one needs to have a cluster with approximately 350 Pt atoms. This would correspond to a total of about 270 Fe atoms. We note that other applications of FePt nanoparticles, such as contrast agents for MRI do not have such strict stability requirements and could be done with much smaller particles, indeed with superparamagnetic nanoparticles.[58]

The second topic in the thermal behavior of these clusters is to estimate how the magnetization would change in these samples as a function of temperature, M(T). Dealing with this problem exactly is difficult, in part because there could be multiple exchange constants within the nanoparticle, and these are not known. Instead, we only want to provide a simple estimate of M(T) and see how this could vary depending on the size and shape of the nanoparticle, and depending on the position of the Fe atoms within the nanoparticle.

The thermal averaged magnitude of a spin in an effective magnetic field is given by

$$\langle S \rangle = S B_S(x) \tag{5}$$

where x is the ratio of the magnetic energy to the thermal energy

$$x = \frac{g \mu_B S H_{eff}}{k_B T} \tag{6}$$

Here $g$ is the gyromagnetic ratio, S is the spin, $H_{eff}$ is the effective magnetic field, and $\mu_B$ is the magnetic moment of the atom. The function $B_S(x)$ is the Brillouin function, and it is given by

$$B_S(x) = \left(\frac{(2S+1)}{2S}\right) coth\left(\frac{(2S+1)x}{2S}\right) - \left(\frac{1}{2S}\right) coth\left(\frac{x}{2S}\right) \tag{7}$$

For bulk Fe with a body centered cubic structure there are 8 nearest neighboring atoms of each Fe atom, so the effective field can be written as $H_{eff} = 8J\langle S \rangle$ where, $J$ is the exchange constant and <S> is the thermal averaged magnitude of the spin.

We can use the value of the critical temperature, $T_c$, to obtain the exchange constant. For small values of $x$ (appropriate near T = $T_c$) the Brillouin function in equation (7) may be expanded to give

$$B(x) = \left(\frac{S+1}{3S}\right) x \tag{8}$$

Substituting this in equation (5) we can get the relation between $J$ and $T_c$

$$J = \frac{3kT_c}{8g\mu_B S(S+1)} \tag{9}$$

Using the value of $T_c$ = 1043 K, appropriate for bulk Fe, we can get an estimate for the exchange constants between nearest neighbor Fe atoms.

In the nanoparticle structures, the effective field acted on each atom is different, so one cannot do standard mean-field theory. Instead we use a simple self-consistent local-mean-field magnetic model.[59] As an example, we show the key elements of the calculation for the $Fe_{20}Pt_{18}$ configuration. We label each Fe atom in our nanoparticles with a different number, as an example, as seen in Fig. 12.

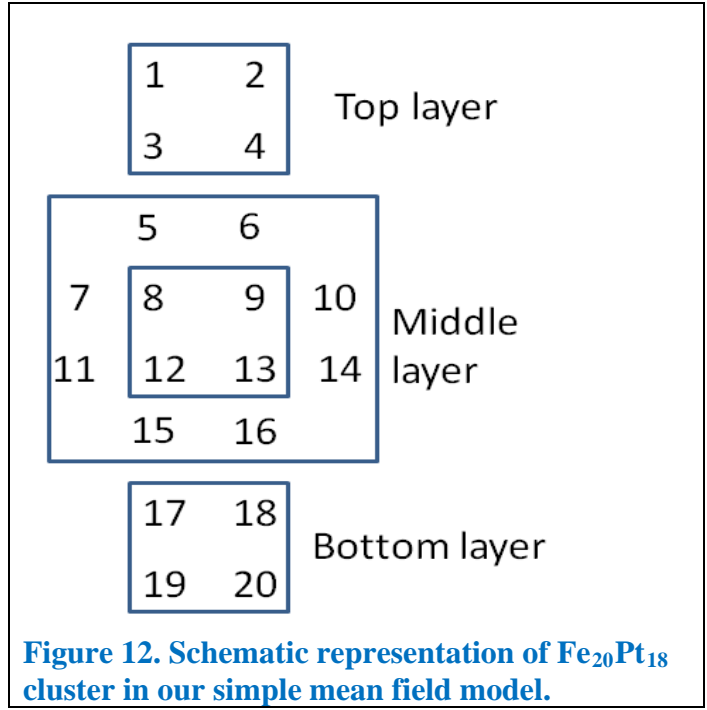

**Figure 12. Schematic representation of $Fe_{20}Pt_{18}$ cluster in our simple mean field model.**

We calculate the effective field acting on each spin, as arising from the exchange coupling from nearby spins. We identify two types of coupling:

1) Coupling of nearest neighbors within a plane with exchange constant $J$
2) Coupling of nearest neighbors between planes, with exchange constant $J_p$

The effective field acting on each atom can be written in terms of $J$ and $J_p$. For example the effective field on spin at site labeled by 1 can be written as

$$H_1 = J(\langle S_2 \rangle + \langle S_3 \rangle) + J_p \langle S_8 \rangle, \tag{10}$$

here we assumed that the spins are all pointing in one direction and $\langle S_n \rangle$ is the thermal averaged magnitude of the spin at site $n$. We can write the effective field equation for each of the site in the cluster, and the thermal averaged magnitude at any site now can be found by using the expression,

$$\langle S_n \rangle = S_n B(x_n) \tag{11}$$

where, $x_n$ in this case can be written as

$$x_n = \frac{g_n \mu_B S_n (\langle S_n \rangle H_n + H_0)}{KT} \tag{12}$$

One then has a set of n coupled equations involving the variables $<S_1>$ to $<S_n>$. We solve this iteratively by picking a site n at random, calculating the effective field at that site and the resulting new value of $<S_n>$. Then a new spin is chosen and the process is repeated until the process converges to final values and all spins have thermal averaged magnitudes which are consistent with the values of the neighboring spins.

We use the following parameters in our calculations $\mu_B = 9.27 \times 10^{-21}$ erg-Gauss, Boltzmann's constant is k = $1.38 \times 10^{-16}$ erg/Kelvin, g = 2, and J = $1.455 \times 10^6$ Oe (bulk values of $J$), S = 1. A small external magnetic field $H_o$ = 100 Oe is used to help orient the moments and speed convergence. To see how the coupling between different planes of Fe atoms affects the results, we use two different values for the perpendicular coupling $J_p$; $J_p = J$ and $J_p = 0.5J$.

Our results for the $Fe_{20}Pt_{18}$ and $Fe_{39}Pt_{40}$ clusters are presented in Fig.13. For both cases the critical temperature is substantially less than the critical temperature of bulk Fe ($T_c = 1043K$). However there is still a substantial moment at room temperature. The reduction in $T_c$ is due to the reduced coordination number of the Fe atoms in the FePt structure. As might be expected, this reduced coordination has a smaller effect for the larger structure because the percentage of atoms at the surfaces and edges is smaller. Indeed we find that the $Fe_{39}Pt_{40}$ has a higher $T_c$ value. We also see that the larger perpendicular coupling case gives a higher critical temperature, as expected. This trend is consistent with experimental data showing that larger FePt nanoparticles have a higher $T_c$.[60]

The thermal average spin values at different sites of the cluster are show in Fig, 13(b), which were done for $J_p$ = 0.5J. As can be easily seen by symmetry, there are only three unique types of site for $Fe_{20}Pt_{18}$ and six unique site for $Fe_{39}Pt_{40}$ cluster. A key result is that even for moderate temperatures, the thermal moments at the different sites throughout the cluster can be quite different. Indeed the thermal averaged values for spins at the outer surfaces and edges can be quite small compared to those in the center for some temperatures.

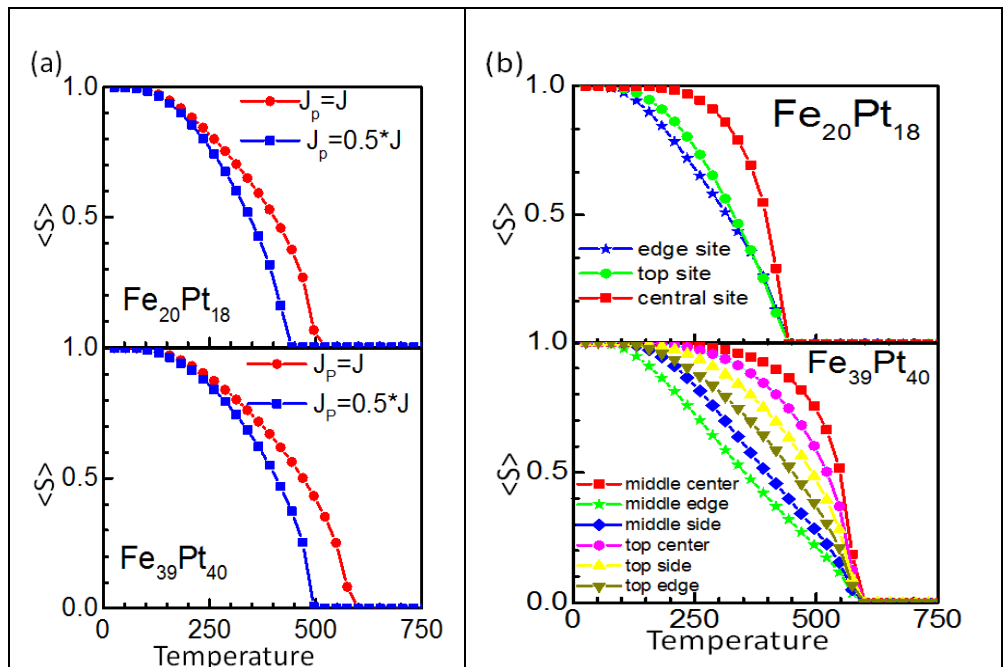

Figure 13. (a) Averaged spin with respect to temperature for $Fe_{20}Pt_{18}$ cluster(on the top) and for $Fe_{39}Pt_{40}$ cluster(on the bottom) (b) averaged spin for inequivalent sites in $Fe_{20}Pt_{18}$(on the top) and $Fe_{39}Pt_{40}$ cluster(on the bottom). The calculation in (b) is done with $J_p = 0.5\ J$

We note that these calculations neglect any magnetic moments in the Pt atoms, and assume that all the Fe exchange values are the same. This is clearly a simplification, but the results presented here, nonetheless, should give some idea of possible thermal behaviors.

## V. Conclusions

We have systematically studied the magnetic properties of FePt $L1_0$ nanoparticles as a function of particle sizes (30, 38, 71, 79, 114, 132, 140, 230, 386 and 484 atoms) and compositions (i.e., consisting of pure Fe and Pt atoms and of alternating planes of Fe and Pt atoms). We find that nanoparticles have much higher magnetic moments than do bulk atoms. This is due to the fact that the MCA arises from the orbital moment coupled with the spin moment and that in the bulk the system orbital moments are almost quenched, whereas in small clusters the orbital moments of the system's atoms are considerably enhanced. We propose that this explains why it is that (as earlier studies have shown) the hybridization of the 5d(Pt) orbitals with 3d(Fe) orbitals produces a high magnetic anisotropy for layered FePt nanoparticles. We also find that clusters with Pt atoms as the central layer have much larger anisotropy than those in which the central layer consists of Fe atoms. The explanation for this is that the central layer has more atoms than other layers in the cluster, and when *these* atoms are of high orbital moment – Pt atoms hybridizing with the Fe atoms below and above – the system as a whole exhibits higher anisotropy than when is central layer consists of Fe atoms, whose orbital moment, in hybridizing with the Pt atoms above and below, is markedly lower. In contrast to the cuboctahedral case [18, 25] bi-pyramidal nanoparticles possess (similar magnitude) MCA mostly at the (large) central Pt layer. This fact may eliminate the need to cap them in order to preserve MCA. Our calculation show that five-layered nanoparticles with approximately 700 atoms can be expected to be useful in magnetic recording applications at room temperatures. Meanwhile, a deeper analysis of the electronic structure of these and other TM nano systems could contribute further to this end

Generally speaking, the relation between the structure and MCA is not yet completely understood for FePt and other binary alloys. For example, as an alternative type of system, 147-atom cuboctahedral FePt clusters encapsulated into Cu-, Au-, and Al matrices were studied theoretically in Ref.[25] where it was found, for example, that surface atoms have larger MCA. Another consideration is the particular role of electron-electron correlation in these systems. This was found to be important for small Fe[61] and FePt[62] clusters and invite further study.

ACKNOWLEDGMENTS

We would like to thank Lyman Baker for critical reading of the manuscript and numerous enlightening comments. The work was supported in part by DOE Grant DE-FG02-07ER46354. REC would like to thank UCF for partial sabbatical support.

.